\documentclass[aps,prb,longbibliography,amsmath,amssymb,superscriptaddress,floatfix,twocolumn]{revtex4-2}

\usepackage{mathrsfs}
\usepackage{physics}
\usepackage{graphicx}
\usepackage[english]{babel}

\graphicspath{{img/}{pict/}{}}

\usepackage{bm}

\usepackage{hyperref}
\hypersetup{pdfstartview={FitH},pdfpagemode={UseNone},
            colorlinks,linkcolor=blue, citecolor=blue, urlcolor=blue,
            bookmarksopen=true, pdfnewwindow=true}

\usepackage[all]{hypcap}

\begin{document}

\title{Enhanced generation of non-degenerate photon-pairs in nonlinear metasurfaces}

\author{Matthew Parry}
\email{Matthew.Parry@anu.edu.au}
\affiliation{Research School of Physics, Australian National University, Canberra, ACT 2601, Australia}
\affiliation{ARC Centre of Excellence for Transformative Meta-Optical Systems (TMOS), Australia}

\author{Andrea Mazzanti}
\affiliation{Dipartimento di Fisica, Politecnico di Milano, Milan, Italy}
\affiliation{Research School of Physics, Australian National University, Canberra, ACT 2601, Australia}

\author{Alexander~Poddubny}
\affiliation{Research School of Physics, Australian National University, Canberra, ACT 2601, Australia}
\affiliation{ITMO University, 49 Kronverksky Pr., Saint Petersburg 197101, Russia}
\affiliation{Ioffe  Institute, Saint Petersburg 194021, Russia}

\author{Giuseppe~Della~Valle}
\affiliation{Dipartimento di Fisica, Politecnico di Milano, Milan, Italy}
\affiliation{Istituto di Fotonica e Nanotecnologie, Consiglio Nazionale delle Ricerche, Milan, Italy}

\author{Dragomir N. Neshev}
\affiliation{Research School of Physics, Australian National University, Canberra, ACT 2601, Australia}
\affiliation{ARC Centre of Excellence for Transformative Meta-Optical Systems (TMOS), Australia}

\author{Andrey A. Sukhorukov}
\email{Andrey.Sukhorukov@anu.edu.au}
\affiliation{Research School of Physics, Australian National University, Canberra, ACT 2601, Australia}
\affiliation{ARC Centre of Excellence for Transformative Meta-Optical Systems (TMOS), Australia}


\begin{abstract}
We reveal a novel regime of photon-pair generation driven by the interplay of multiple bound states in the continuum resonances in nonlinear metasurfaces. 
This non-degenerate photon-pair generation
is derived from the hyperbolic topology of the transverse phase-matching and can enable orders-of-magnitude enhancement of the photon rate and spectral brightness, as compared to the degenerate regime. We show that the entanglement of the photon-pairs can be tuned by varying the pump polarization, which can underpin future advances and applications of ultra-compact quantum light sources.
\end{abstract}

\maketitle

\begin{figure*}[t]
    \centering
    \includegraphics[width=0.8\textwidth]{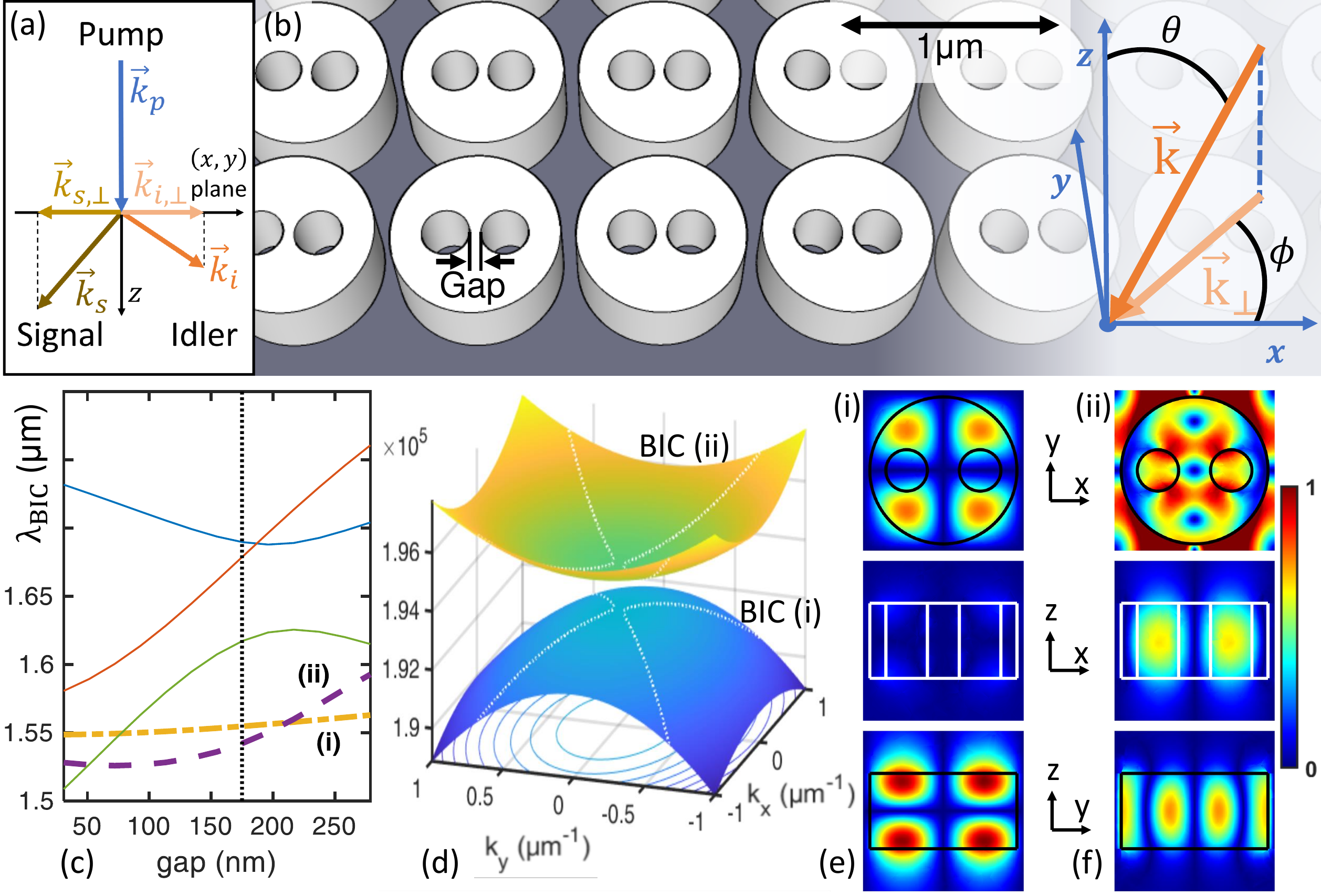}
	\caption{The metasurface and its modes. 
	(a)~Diagram of transverse phase matching.
	(b)~Metasurface design and the coordinate axes: 
	\(\theta\) is the polar and \(\phi\) the azimuthal angle. 
	(c) BIC wavelengths vs. the gap between the two holes.
	(d) The dispersion for two quasi-BICs: (i) lower and (ii) upper surfaces.  Dotted white lines: two-BIC phase matching for \(\lambda_p=774.22\)~nm.  (e) \& (f) Fields \(|\vb{E}|\) of the two BICs.}
    \label{fig:1}
\end{figure*}

Metasurfaces (MSs) offer an ultra-compact and versatile platform for enhancing nonlinear optical processes, including harmonic generation and frequency mixing~\cite{Li:2017-17010:NRM, DeAngelis:2020:NonlinearMetaOptics}.  To achieve such nonlinear interactions
in bulk crystals and waveguides one requires extended propagation distances, but in metasurfaces a strong enhancement of light-matter interactions can be achieved with sub-wavelength thicknesses through the excitation of high-quality factor optical resonances.  Notably, this can be achieved by designing Bound State in the Continuum (BIC) resonances~\cite{Hsu:2016-16048:NRM, Rybin:2017-243901:PRL, Koshelev:2018-193903:PRL, Koshelev:2020-288:SCI}, which facilitate a high confinement of the optical field within the nonlinear material~\cite{Vabishchevich:2018-1685:ACSP, Carletti:2019-33391:OE, Huang:2021-:NANT}.

In addition to classical frequency mixing, nonlinear metasurfaces can also, through spontaneous parametric down-conversion (SPDC), generate entangled photons with a strong degree of spatial coherence~\cite{Poddubny:2020-147:QuantumNonlinear}. 
SPDC in carefully engineered metasurfaces has the potential to drive fundamental advances in the field of ultra-compact multi-photon sources~\cite{Solntsev:2021-327:NPHOT} that operate at room temperature, and are also suitable for integration in end-user devices with applications that include quantum imaging~\cite{Basset:2019-1900097:LPR} and free-space communications~\cite{Steinlechner:2017-15971:NCOM}. 
Traditionally, SPDC is performed in bulk nonlinear crystals with dimensions up to centimeters in length, while integrated waveguides have enabled a reduction of the footprint to millimeter~\cite{Orieux:2013-160502:PRL} and down to 
\(100~\mu m\) length-scales~\cite{Collins:2013-2582:NCOM}.
At the sub-wavelength scale, generation of photon-pairs was reported experimentally from a single AlGaAs nano-resonator~\cite{Marino:2019-1416:OPT}, lithium niobate metasurfaces~\cite{Santiago-Cruz:2103.08524:ARXIV}, and studies were also conducted on monolayers of MoS\textsubscript{2}~\cite{Saleh:2018-3862:SRP} and carbon nano-tubes~\cite{Lee:2017-1605978:ADM}.

Importantly, SPDC in ultra-thin nonlinear layers~\cite{Okoth:2019-263602:PRL, Okoth:2020-11801:PRA, Santiago-Cruz:2021-653:OL} can give rise to strong spatial correlations and allow quantum state engineering without the constraints of longitudinal phase-matching.
It has been proposed that a so-called `accidental' BIC at the pump frequency can increase the photon rate at a single nanoresonator~\cite{Poddubny:1808.04811:ARXIV}, while a photonic crystal slab with a BIC resonance can enhance SPDC in a monolayer of WS\textsubscript{2}~\cite{Wang:2019-341:PRJ}, although the theoretically estimated rate was still much lower than with conventional sources.
There is now a strong interest in new concepts and practical approaches for even stronger enhancement of the brightness of SPDC photon-pair generation in sub-wavelength scale structures.

In this paper, we present a general approach for orders-of-magnitude enhancement of the photon-pair generation rate and spectral brightness in nonlinear metasurfaces. We reveal, for the first time to our knowledge, that non-degenerate SPDC efficiency can be dramatically increased when the signal and idler photons are supported by BIC resonances at different frequencies.
We demonstrate how these features can be realized in practice by engineering the symmetry of the metasurface to deliver a projected spectral brightness five orders of magnitude higher than for an unpatterned film. 
Our results are fundamentally different from the recently demonstrated SPDC generation using linear metasurface-lenses that engineers the quantum state by focusing the pump at multiple spots in a bulk crystal~\cite{Li:2020-1487:SCI}.

We demonstrate that a metasurface with reduced global rotational symmetry allows much greater flexibility in tailoring the dispersion from several BICs at which the signal and idler photons are generated. Consequently, we can control the form of the transverse phase matching for SPDC according to the energy and momentum conservation as illustrated in Fig.~\ref{fig:1}(a), 
%
\begin{eqnarray}
    \label{eqn:phase-matching-k}
    \vb{k}_{p,\perp} &=& \vb{k}_{s,\perp} + \vb{k}_{i,\perp}, \\
    \label{eqn:phase-matching-freq}
    \omega_p(\vb{k}_{p,\perp}) &=& \omega_s(\vb{k}_{s,\perp})+\omega_i(\vb{k}_{i,\perp}),
\end{eqnarray}
%
where the indices refer to pump (p), signal (s), and idler (i) photons with the corresponding frequencies $\omega$. The wave-vectors $\vb{k}$ define the propagation directions in free space, and $\perp$ indicates the transverse components in the plane of the metasurface.

We focus on a metasurface with \(D_{2h}\) symmetry and outline other possibilities in Supplementary~S2. According to our concept, we designed a metasurface composed of a square array of cylinders with two holes, or \emph{Ghost Oligomers}, inserted to remove the 90$^\circ$ rotational symmetry [Fig.~\ref{fig:1}(b)]. This structure supports multiple extended photonic-crystal like BICs~\cite{Hsu:2016-16048:NRM}, where the field localization arises from a mismatch between the symmetry of the collective modes and the available radiation channels.

We consider resonators made of Al\textsubscript{0.2}Ga\textsubscript{0.8}As, which possesses strong quadratic nonlinearity and can be manufactured with established procedures. We chose a (111) crystal orientation as it provides the best off-BIC conversion efficiency in the normal propagation direction (Supplementary~S3) and hence gives a better measure of the enhancement due to the BIC.
In our modeling we have omitted a substrate to focus on the generic features. Adding a substrate will convert the BICs to quasi-BICs, due to the up/down asymmetry introduced.  But, as we discuss in the following, the photon generation is associated with quasi-BICs formed by off-normal angles, thus confirming that the ideal BICs with formally infinite quality factors are not required. We therefore conclude that a strong enhancement will still be achieved when a substrate is present.
Our general approach can also be applied to metasurfaces made of different nonlinear materials, including Lithium Niobate where efficient classical frequency conversion has been demonstrated~\cite{Ma:2002.06594:ARXIV, Huang:2019-69:NANM, Fedotova:2020-8608:NANL}.

An important feature of BICs is that changes in the dimensions of the meta-atoms will not destroy a photonic-crystal like BIC so long as the relevant symmetry is maintained.  This enables simultaneous tuning of the resonant frequencies for several BICs by adjusting the design parameters, such as the separation between the hole pairs as shown in Fig.~\ref{fig:1}(c).
By shifting the position of the holes the proportion of the electric field within these two low refractive index regions changes, which in turn changes the energy of the modes. The way in which each mode is affected depends on the profile of the mode over the region through which the holes move. Similar configurability is found with the other dimensions of the meta-atoms.
This tunability offers an important advantage with SPDC generation when compared to using so-called `accidental' BICs~\cite{Carletti:2018-33903:PRL} in individual nano-resonators which only appear at very specific resonator dimensions.

The dispersion of the two BICs studied in this paper are shown in Fig.~\ref{fig:1}(d) which, importantly, have opposite dispersion.
Their mode profiles are shown in Figs.~\ref{fig:1}(e) \&~(f), where the fields are normalized to the peak value inside the resonator for each BIC.  They are also marked (i) and (ii) in Fig.~\ref{fig:1}(c), where the dotted line shows the value of the gap used in the two BIC study.  The single BIC study uses BIC (i) [Fig.~\ref{fig:1}(e)] with a gap of \(52\)~nm, which gives a good separation from the other BICs.

\begin{figure}[t]
    \centering
    \includegraphics[width=\columnwidth]{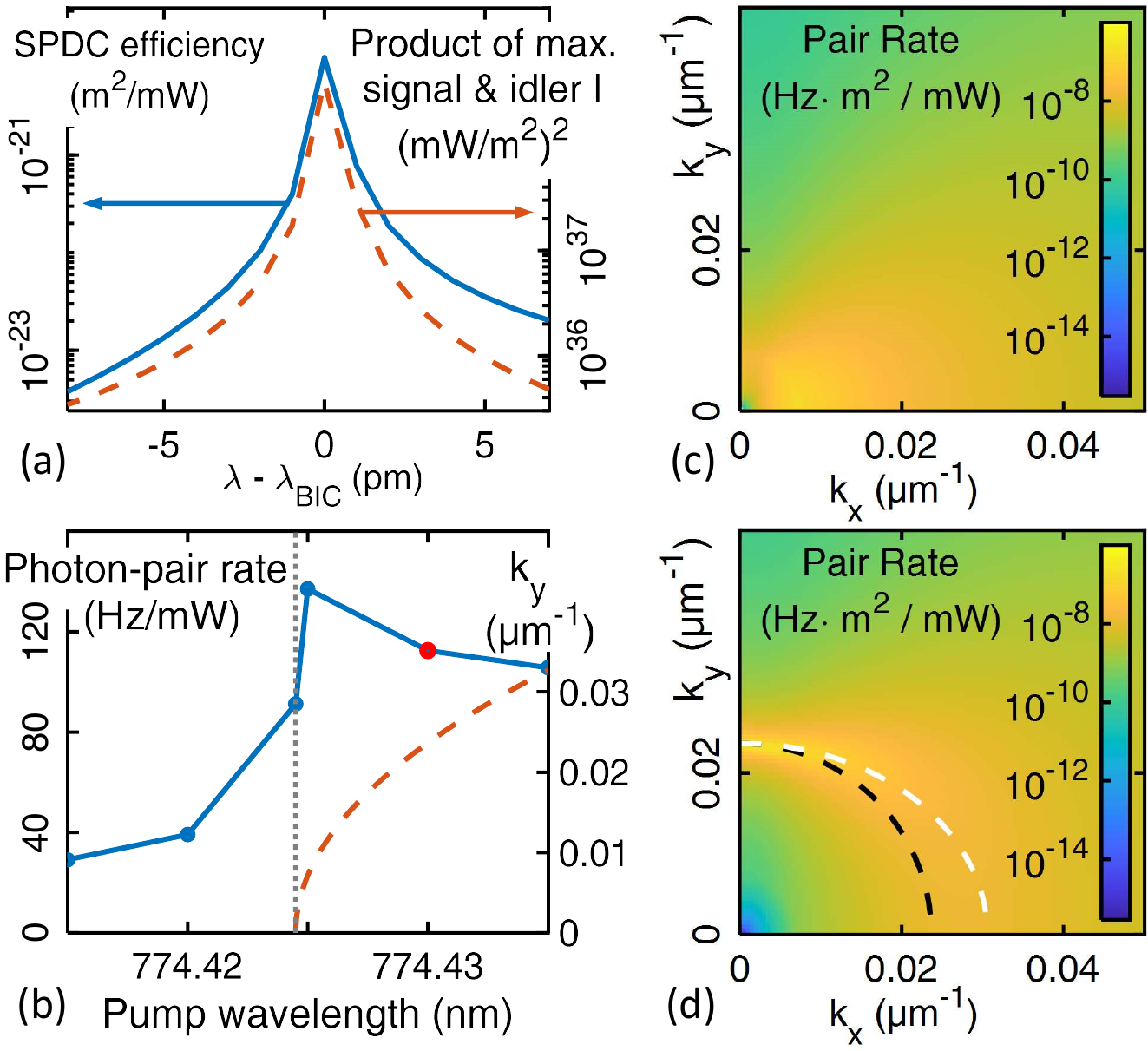}
	\caption{SPDC at a single BIC. 
	(a) Solid line: the generation rate for a horizontally polarized, normally incident pump, \(\ket{HH}\) signal and idler state, \(\phi=79^\circ\) and  \(\theta_s=0.2^\circ\) for the signal, opposite angle for the idler. 
	Dashed: The product of the maximum intensity inside the resonator of the signal and idler fields. 
	(b) Solid line: generation rate vs. the pump wavelength. The red dot marks the wavelength used in (d) and Fig.~\ref{fig:3}.  
	Dashed: \(k_y\) at phase matching when \(k_x=0\). (c,d) Generation rate in $k$-space for (c)~phase matching at the \(\Gamma\) point and (d)~for \(\lambda_p=774.43\)~nm.  White dashed line: phase matching condition;  Black dashed line: the path of constant \(\theta_s\) in Fig.~\ref{fig:3}(a).
    }
    \label{fig:2}
\end{figure}

As a first study we present an analysis of SPDC with the signal and idler photons being generated at a single BIC. 
We calculated the SPDC generation rate via the quantum-classical correspondence between SPDC and Sum Frequency Generation (SFG)~\cite{Poddubny:2016-123901:PRL, Lenzini:2018-17143:LSA, Marino:2019-1416:OPT, Poddubny:2020-147:QuantumNonlinear}, 
which is exact in the absence of other nonlinear effects. We performed full SFG numerical simulations, and used these results to predict the efficiency of quantum photon-pair generation through SPDC,
\begin{equation}
\Xi_{\scriptscriptstyle SPDC} =\frac{1}{\qty(2\pi)^3}\frac{\lambda_p^2}{\lambda_s^{\scriptscriptstyle BIC} \lambda_i^{\scriptscriptstyle BIC}} \frac{\Phi_{\scriptscriptstyle SFG}}{\Phi_s\Phi_i},
\end{equation}
where \(\Phi_{\scriptscriptstyle SFG}\) is the zeroth order farfield SFG intensity, and \(\Phi_s\) and \(\Phi_i\) are the incident signal and idler intensities.  For SPDC calculations we take the small angle approximation that in the farfield the $z$ component is zero, so that \(\ket{H}\) and \(\ket{V}\) notations indicate the polarization primarily along the $x$- and $y$-axis, respectively.
The solid line in Fig.~\ref{fig:2}(a) shows the results of such modeling. 

\begin{figure}[t]
    \centering
    \includegraphics[width=\columnwidth]{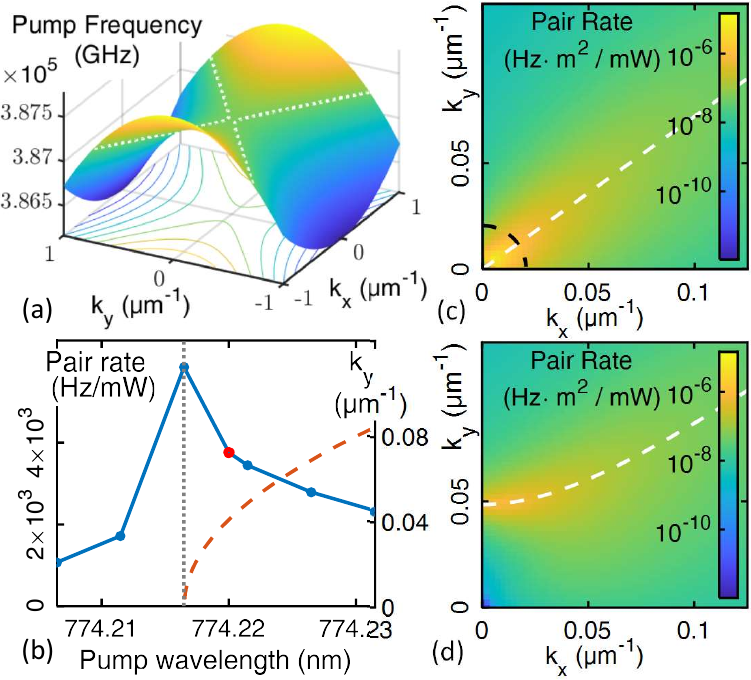}
	\caption{Non-degenerate signal and idler at different BICs.  (a)
	Transverse phase matching condition according to Eq.~(\ref{eqn:phase-matching-freq}).  The dotted white line 
	corresponds to 
	\(\lambda_p=774.22\)~nm.  
	(b) Solid: generation rate vs. the pump wavelength. The red circle is the integration of~(d). 
	Dashed: \(k_y\) at transverse phase matching when \(k_x=0\).  
	(c,d) Generation rate in $k$-space at (c)~phase matching at the \(\Gamma\) point (\(\lambda_p=774.2165\)~nm) and (d)~at \(\lambda_p=774.22\)~nm. White dashed line: phase matching condition;  Black dashed line: the path of constant \(\theta_s\) in Fig.~\ref{fig:3}(c).
	}
    \label{fig:4}
\end{figure}

We then confirmed that the SPDC generation is approximately proportional to the product of the maximum intensity of the signal and idler electric fields inside the resonator. We present a 
typical
dependence of this quantity near a BIC resonance, obtained from linear simulations,  with the dashed line in Fig.~\ref{fig:2}(a), where there is a slightly off-normal signal and idler (\({\theta_s=0.2^\circ}\)) and the pump wavelength is \(\lambda_p=775\)~nm. 
We can see that the shapes of solid and dashed lines match closely.
This is a physically important observation, as it means that the designed metasurface optimally translates the BIC enhancement of the signal and idler to a corresponding increase of SPDC. 

We conducted studies of the intensity enhancement for different angles of \(\theta_s\) and \(\phi_s\).  The simulated intensities are fitted to a Lorentzian function \(L(\omega,\vb{k})\) (Supplementary~S5) and from this fitted function we calculate
\begin{equation}\label{eqn:SPDCfit}
 \begin{split}
\Xi_{\scriptscriptstyle SPDC} &=\frac{1}{\qty(2\pi)^3} \frac{\lambda_p^2}{\lambda_s^{\scriptscriptstyle BIC} \lambda_i^{\scriptscriptstyle BIC}}\\
&\times \sum_{\genfrac{}{}{0pt}{3}{\ket{HH},\ket{HV},}{\ket{VH},\ket{VV}}} \Xi_0L(\omega_s,\vb{k_s})L(\omega_p-\omega_s,\vb{k}_p-\vb{k_s}),
 \end{split}
\end{equation}
where \(\Xi_0\) is the ratio of the SFG zeroth diffraction order farfield intensity to the product of the maximum signal and idler intensities inside the resonator (Supplementary~S4).  The sum is taken over all polarization combinations of horizontal (\(\ket{H}\)) and vertical (\(\ket{V}\)) for the signal and idler, which forms the polarization basis for our wavefunction.  The total photon pair generation rate across all angles and wavelengths can then be calculated via
\begin{equation}\label{eqn:pair-rate}
\frac{1}{A_{tot}\Phi_p}\dv{N_{pair}}{t} =  
\iiint\Xi_{\scriptscriptstyle SPDC} 
\dd{\omega_s}\dd{k_{s,x}}\dd{k_{s,y}},
\end{equation}
where \(A_{tot}\) is the total sample area and \(\Phi_p\) the incident pump intensity.

\begin{figure}[t]
    \centering
    \includegraphics[width=\columnwidth]{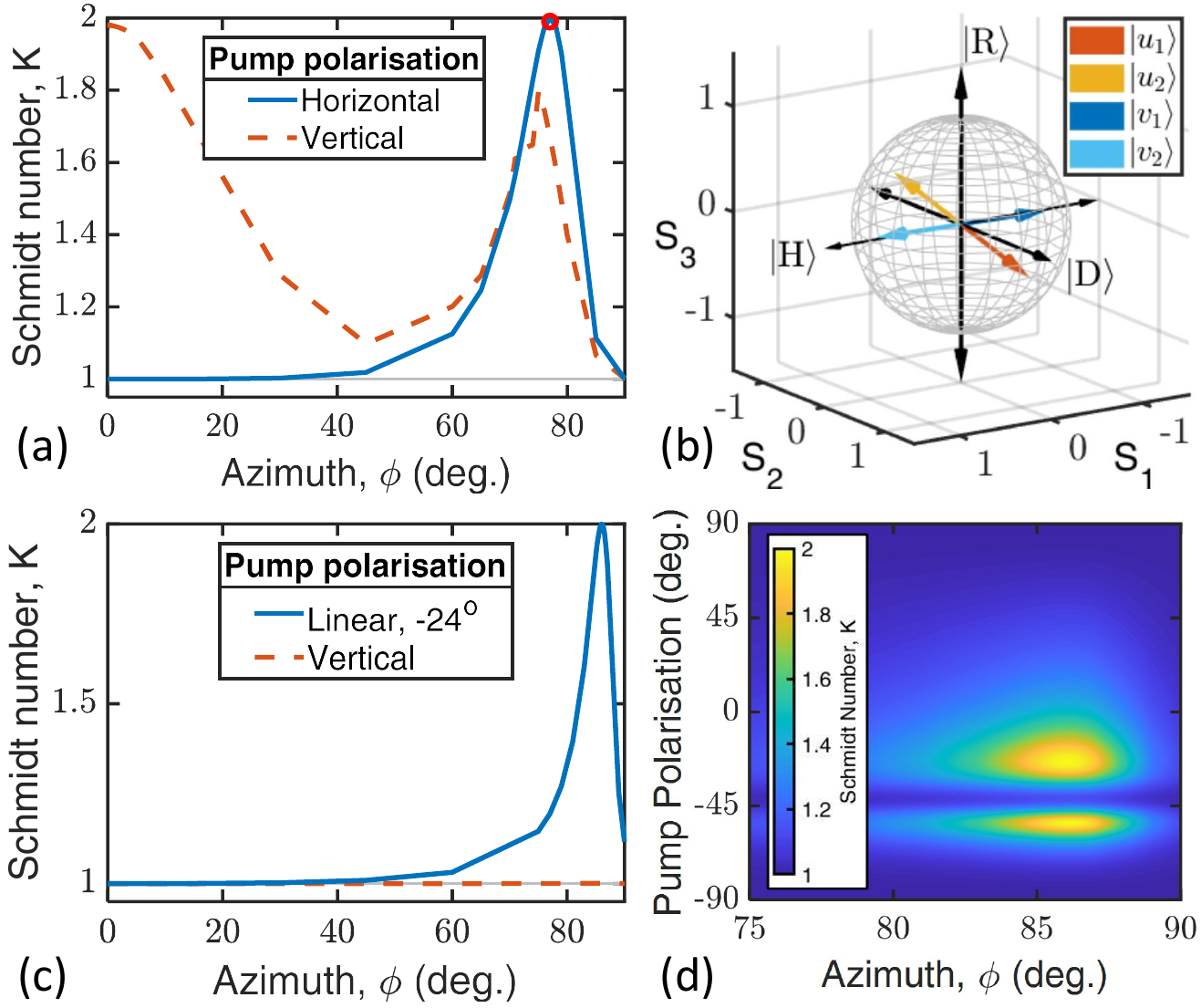}
	\caption{Entanglement of the signal and idler photons generated from: (a) \& (b) a single BIC with \(\lambda_p=774.43\)~nm; (c) \& (d) two different BICs with \(\lambda_p=774.2165\)~nm. (a,c) Schmidt number vs. the photon angle.
	(b) Schmidt decomposition of the wavefunction plotted on the Poincar\'e sphere for the red dot in~(a).
	(d)~Schmidt number vs. the pump polarization and photon angle.
	}
    \label{fig:3}
\end{figure}

If we only integrate Eq.~(\ref{eqn:pair-rate}) over frequency then we will obtain the pair-rate as a function of the signal wavevector, as seen in Figs.~\ref{fig:2}(c) \&~(d).  Fig.~\ref{fig:2}(c) has a pump wavelength at which the transverse phase matching condition occurs at the \(\Gamma\) point and Fig.~\ref{fig:2}(d) has \(\lambda_p=774.43\)~nm, which was chosen for illustrative purposes.  Note that the pair-rate falls off near the \(\Gamma\) point as symmetry protected BICs have singularities at the \(\Gamma\) point.  We therefore only see enhancement in the off-\(\Gamma\), quasi-BIC, regime.  In Fig.~\ref{fig:2}(d) the dashed white line shows the angle in $k$-space at which the transverse phase matching condition occurs, which matches the peak in generation as expected.

By then integrating over $k$-space we obtain the total photon-pair generation rate as a function of the pump wavelength, as shown with a solid line in Fig.~\ref{fig:2}(b).  Notably, the peak does not occur at the \(\Gamma\) point but just beyond it, as can be seen by comparing this plot with the dashed line showing the value of \(k_y\) at the transverse phase matching condition when \(k_x=0\).  This, again, is due to the singularity at the \(\Gamma\) point.  We calculate the theoretical peak brightness of this quasi-BIC to be 110 Hz/mW/nm over a 1.3~nm bandwidth, which is \(2\times10^3\) that of simulations of an unpatterned nonlinear film.

Next, we studied the case where the signal and idler are generated at two different BICs.  
The most distinguishing feature here is the occurrence of hyperbolic transverse phase matching as shown in Fig.~\ref{fig:4}(a), calculated according to the right-hand side of Eq.(\ref{eqn:phase-matching-freq}) for the normally incident pump with $\vb{k}_{p,\perp}=0$.
The condition for the pump wavelength of \(\lambda_p=774.22\)~nm is shown by the dotted white line in Fig.~\ref{fig:1}(d), Figs.~\ref{fig:4}(a) and~(d).   

For the two BIC case, in Fig.~\ref{fig:4}(c) we show the angular dependence in $k$-space of the photon-pair generation when transverse phase matching occurs at the \(\Gamma\) point, and in Fig.~\ref{fig:4}(d) the case in which \(\lambda_p=774.22\)~nm.  As before there is a peak in generation at the transverse phase matching condition (dashed white line).  In Fig.~\ref{fig:4}(b) we see that the photon-pair generation rate is almost two orders of magnitude higher than for the single BIC case in Fig.~\ref{fig:2}(b).  A factor of $6$ of the enhancement can be attributed to different mode profiles and their overlap (Supplementary~S4). Importantly, an order-of-magnitude increase is due to hyperbolic phase matching, whereby SPDC enhancement occurs for a much broader range of transverse photon wavevectors, in contrast to single-BIC case with elliptical phase-matching only allowing a small range of wavevectors close to the $\Gamma$ point.
We calculated the theoretical peak brightness to be 4900 Hz/mW/nm over a 1.2~nm bandwidth, which is \(10^5\) that of simulations for an unpatterned nonlinear film. Such predicted brightness enhancement is also much stronger then for metasurfaces based on Mie-like resonances~\cite{Santiago-Cruz:2103.08524:ARXIV}.

The two elliptic paraboloids in Fig.~\ref{fig:1}(d) can only sum to a hyperbolic paraboloid because the metasurface lacks \(\flatfrac{\pi}{2}\) in-plane rotational symmetry.
The \(D_{2h}\) symmetry of the MS means that the BICs must have an elliptic paraboloid dispersion as that matches the rotation and mirror symmetries of the MS.  In our case \(k_x\) and \(k_y\) are the major and minor axes, but a set of axes can always be chosen such that the dispersion is of the form
\begin{equation} \label{eq:dispersionParabolic}
\omega(\vb{k}_{\perp}) = \zeta\left(\frac{k_x^2}{a^2} + \frac{k_y^2}{b^2}\right) + \omega_0,
\end{equation}
where \(\zeta=\pm1\).  The transverse phase matching condition, for a pump with normal incidence, is thus given by
\begin{equation}\label{eqn:hyperbolic}
 \begin{split}
\omega_p&\left(\vb{k}_{p,\perp}=0\right) = \omega_s\left(\vb{k}_{s,\perp}\right) + \omega_i\left(\vb{k}_{i,\perp}=-\vb{k}_{s,\perp}\right)\\
=& k^{2}_{s,x}\left(\frac{\zeta_s}{a_s^2}+\frac{\zeta_i}{a_i^2}\right) +  k^{2}_{s,y}\left(\frac{\zeta_s}{b_s^2}+\frac{\zeta_i}{b_i^2}\right )
+ \omega_{s,0}+\omega_{i,0}\\ 
=&\zeta_x\frac{k^{2}_{s,x}}{a_p^2} + \zeta_y\frac{k^{2}_{s,y}}{b_p^2} + \omega_{p,0},
 \end{split}
\end{equation}
which can be either an elliptic or hyperbolic paraboloid, with the latter enabling enhanced photon-pair generation as discussed above.
In contrast, for a MS with \(D_{4h}\) symmetry (such as a slab with a square array of single holes) \(\zeta_x\equiv \zeta_y\) and \(a_p\equiv b_p\) which excludes a hyperbolic type of transverse phase matching. 

We determined the polarization entanglement of the generated signal and idler by performing
a Schmidt decomposition (Supplementary~S7), where the Schmidt number
\(K=1\) indicates no entanglement and \(K=2\) maximum entanglement.  In Fig.~\ref{fig:3}(a) we show that for the single BIC case the entanglement peaks at \(K=2\) when the azimuthal angle of the signal photon emission is \(\phi_s=77^\circ\) and the pump is horizontally polarized.  The
maximally entangled signal and idler pair state \(\{\ket{u_1},\ket{v_1}\}\) or \(\{\ket{u_2},\ket{v_2}\}\) are visualized in Fig.~\ref{fig:3}(b).
For the two BIC case we can switch from full to no entanglement by changing the linear polarization of the pump, see Figs.~\ref{fig:3}(c) and~(d).  

In conclusion, we have developed a novel method of enhanced photon-pair generation via symmetry protected BICs in nonlinear metasurfaces, in which symmetry is chosen to realize a hyperbolic transverse phase matching condition. In terms of the brightness of the photon pair-generation, our nanoscale platform provides five orders of magnitude improvement over unpatterned films. Additional benefits are the tunability of the photon wavelengths and the degree of polarization entanglement.
We anticipate that these predictions can stimulate significant experimental advances in 
miniaturized quantum light sources based on ultrathin nonlinear metasurfaces for fundamental research and applications.

\begin{acknowledgments}
We acknowledge the support by the Australian Research Council (DP190101559, CE200100010).
\end{acknowledgments}


\bibliography{bibliography}

\end{document}